\documentclass[10pt,twocolumn,twoside,letterpaper]{IEEEtran}

\usepackage{geometry}
\usepackage{url}

\usepackage{cite}
\usepackage{amsmath,amssymb,amsfonts}
\usepackage{graphicx}
\usepackage{textcomp,nicefrac}
\usepackage{subcaption}
\usepackage{hhline}
\usepackage[dvipsnames]{xcolor}
\usepackage[final]{changes}

\begin{document}
\title{Simulation of Charge Collection in a Boron-coated Straw Detector for Emerging Fuel Cycles}

\author{Ming Fang,
	   and Angela Di Fulvio
\thanks{Manuscript received December 9, 2022. This work was funded by the STTR-DOE grant DE-SC0020733. We thank Jeff Lacy and Athanasios Athanasiades at PTI for their support to ensure optimal operation of the NMC.}
\thanks{M. Fang, and A. Di Fulvio are with the Department of Nuclear, Plasma, and Radiological Engineering, University of Illinois at Urbana-Champaign, Urbana, IL 61801, United States (telephone: 217-305-1769, e-mail: mingf2@illinois.edu;difulvio@illinois.edu).}}

\maketitle

\pagenumbering{gobble}

\begin{abstract}
Tristructural-isotropic (TRISO) fuel is currently one of the most mature fuel types for candidate advanced reactor types, namely pebble bed reactors (PBRs). In PBRs, TRISO-fueled pebbles can be re-introduced into the core several times before reaching their target burnup. Non-destructive techniques capable of assaying ${}^{235}$U mass in the pebble are therefore needed for nuclear material control and accountability during fuel recirculation. In this work, we have developed a new boron-coated straw (BCS) based neutron multiplicity counter (NMC) to estimate ${}^{235}$U mass in each pebble. BCS detectors are chosen for their inherent high insensitivity to gamma rays that will enable their use to assay also irradiated pebbles and high neutron detection efficiency, comparable to ${}^{3}$He detectors. The BCS-based NMC that we have designed was built by Proportional Technologies, Inc. (PTI) Houston, TX. In this work, we report the system-level simulation of the BCS-based NMC and the straw-level charge collection simulation coupled with a custom software to tally the detected pulse integral from the list mode energy deposited. We have developed a high-fidelity model of the NMC to simulate the response of a single straw detector to a ${}^{252}$Cf source. The simulated die-away time, single neutron count rate, and double neutron count rate agree well with measured values, with a relative difference within $\pm$0.4\%. The simulated charge spectrum agrees well with the measured one in the case of a round straw.  We plan to use the NMC to perform active and passive interrogation of fresh and spent fuel pebbles.
\end{abstract}

\section{Introduction}
\IEEEPARstart{T}{RISO} fuel is currently one of the most mature fuel types for advanced reactor types, namely the PBR. In PBRs, TRISO-fueled pebbles are cycled through the core continuously until they reach a target burnup. Quantifying the ${}^{235}$U mass in TRISO fuel pebbles is crucial for special nuclear material (SNM) accountability and for pebble tracking, when combined with computational methods~\cite{doesttraward}. Hence, knowledge of ${}^{235}$U mass is needed to validate the reactor simulations and for safeguards of SNM.
In this application, two constraints need to be met: high sensitivity to assay a small ${}^{235}$U mass and rejection of gamma rays to assay ${}^{235}$U and fissile fission products such as ${}^{244}$Cm in partially burned fuel. To meet these constraints, we have developed a BCS-based NMC. BCS, unlike ${}^{3}$He, are readily available and cost-effective. They show a faster response than ${}^{3}$He~\cite{lacy2009boron} and are insensitive to gamma rays, unlike organic scintillators, hence enabling the assay of partially spent pebbles. The BCS used in this work features a pie-shaped cross section with increased surface area compared to traditional round-shaped straws. To the best of our knowledge, this is the first NMC based on pie-shaped straws, which provides higher detection efficiency and shorter die-away time than the round straws. \\ In this paper, we will report the most relevant results of a high-fidelity model of the NMC experimentally validated with a ${}^{252}$Cf source at UIUC.

\section{Development and testing of the NMC}
\subsection{Experimental methods}
\begin{figure}[!htbp]
     
    \centering
    \includegraphics[width=\linewidth]{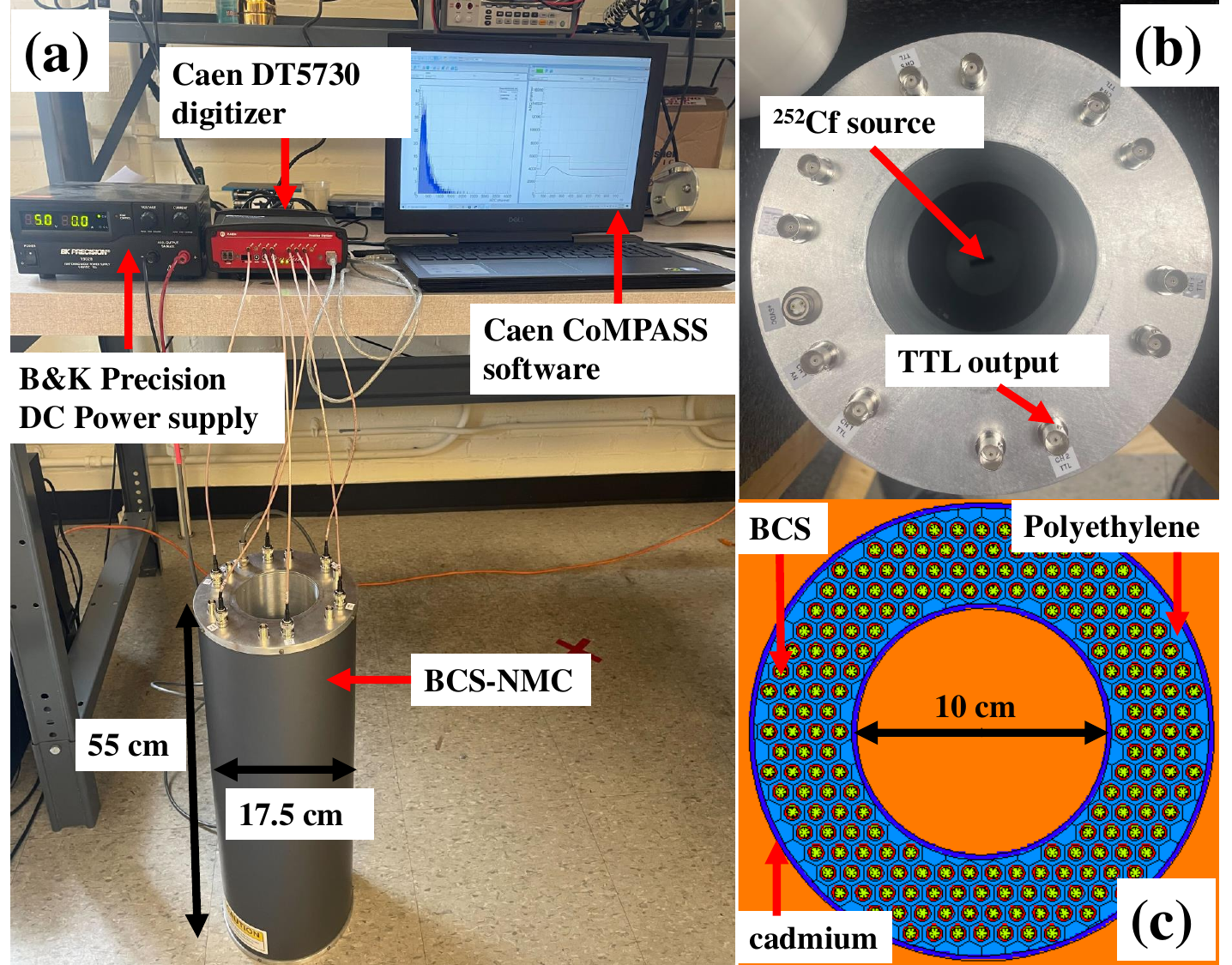}
    \caption{(a): Overview of the experimental setup. (b): Top view of the NMC. (c): Cross-sectional view of the NMC model.}
    \label{fig:cf252 measurement}
    % \vspace{-2em}
\end{figure}

Fig.~\ref{fig:cf252 measurement} shows the BCS-based NMC measuring a 185-kBq (activity date: 2022-02-15) ${}^{252}$Cf source. As shown in Fig.~\ref{fig:cf252 measurement}c, the NMC consists of 192 BCS detectors with 0.9091 cm inter-distance embedded in a high-density polyethylene (HDPE) matrix. The inner and outer radius of the NMC is 5~cm and 8.75 cm, respectively.  The inner and outer surfaces of the NMC are coated with a layer of 0.5~mm-thick cadmium to prevent thermal neutron reentering the sample cavity. Each BCS detector is  40-cm long and has a  4.7-mm outer diameter. A 1.3-$\mu$m thick layer of $\mathrm{{B}_4C}$ (96\% ${}^{10}$B enrichment) is deposited on the inner surface of the straw to absorb the thermal neutrons. As shown in Fig.~\ref{fig:drift-lines}, there are 6 septa in the straw to increase the surface area in the volume, thus increasing the detection efficiency and reducing the system die-away time. A pair of ${}^{7}$Li/alpha ions are produced in the neutron absorption reaction: 
\begin{equation}
    \mathrm{n} + {}^{10}\mathrm{B} \to \begin{cases}
    {}^7\mathrm{Li}(1.01\mathrm{MeV})+{}^4\mathrm{He}(1.71\mathrm{MeV}), &6.3\%\\
    {}^7\mathrm{Li}(0.84\mathrm{MeV})^{*} +{}^4\mathrm{He}(1.47\mathrm{MeV}),              & 93.7\%
\end{cases}
\end{equation}
The straw is filled with Ar/$\mathrm{{CO}_2}$ (9:1) mixture gas at 0.7 atmosphere to detect ${}^{7}$Li/alpha ions. The alpha/Li-7 particles ionize the gas and the ionized electrons drift towards the anode wire under the electrical field in the straw space, which induces electrical signals on the anode wire. The bias voltage applied to the anode wire of the straw is 750V. Signals of 32 straws are bundled together to one TTL output and the six TTL outputs are connected to a 500MS/s DT5730 digitizer~\cite{caencompass} for list mode data acquisition. During the measurement, a voltage threshold of 13~mV is applied in the BCS-based NMC to reject the low-amplitude gamma ray events, resulting in a neutron thresholding efficiency of 92\%~\cite{neutronTE}. We have developed a program to perform real time analysis to calculate the die-away time and coincidence count rates using shift-register algorithms. In this work, we report the results using a signal-trigger shift register protocol~\cite{Croft2012152}.

 \begin{figure*}[ht]
    \centering
    \includegraphics[width=.6\textwidth]{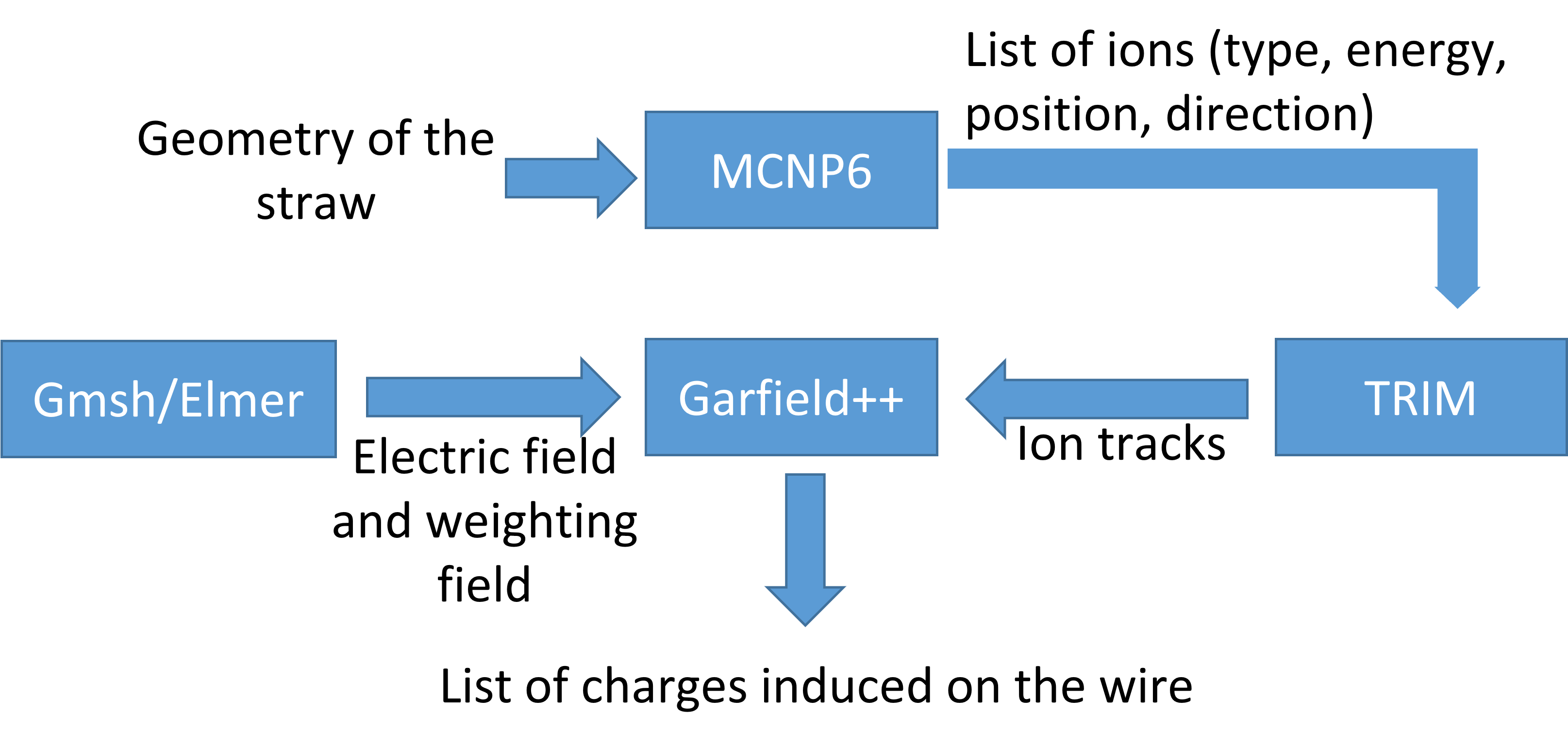}
    \caption{Charge spectrum simulation scheme.}
    \label{fig:workflow}
\end{figure*}
 %\vspace{-2em}
\subsection{Simulation methods}
\subsubsection{System-level model}
We have developed a system -level model in \textit{MCNP 6.2}~\cite{brown2017s}, as shown in Fig.~\ref{fig:cf252 measurement}c. The system model accurately reproduces the geometry of the BCS-based NMC. MCNP's PTRAC card is used to record the energy deposition events by alpha/Li-7 particles in the straw's sensitive volume. A custom-developed software is used to parse the PTRAC output and extract the information of alpha/Li-7 energy deposition events. To account for the neutron count loss due to the voltage threshold, an energy threshold of 57~keV is applied in the simulation.
\subsection{Straw-level model}
\begin{figure}[!htbp]
     
    \begin{subfigure}[t]{\linewidth}
        \centering
        \includegraphics[width=\linewidth]{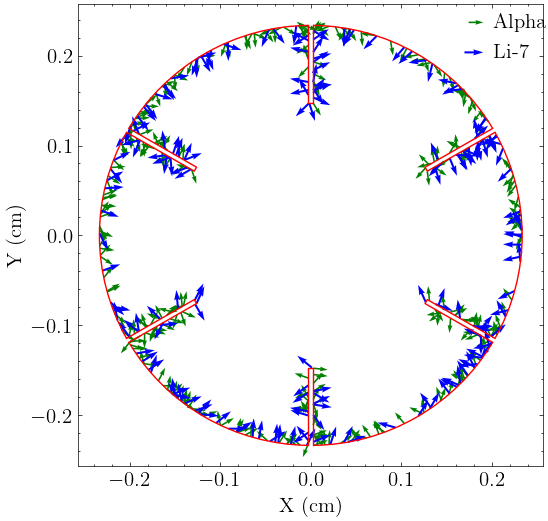}
        \caption{ }
        \label{fig:init-ion-pos}
    \end{subfigure}\hfil
    \begin{subfigure}[t]{\linewidth}
        \centering
        \includegraphics[width=\linewidth]{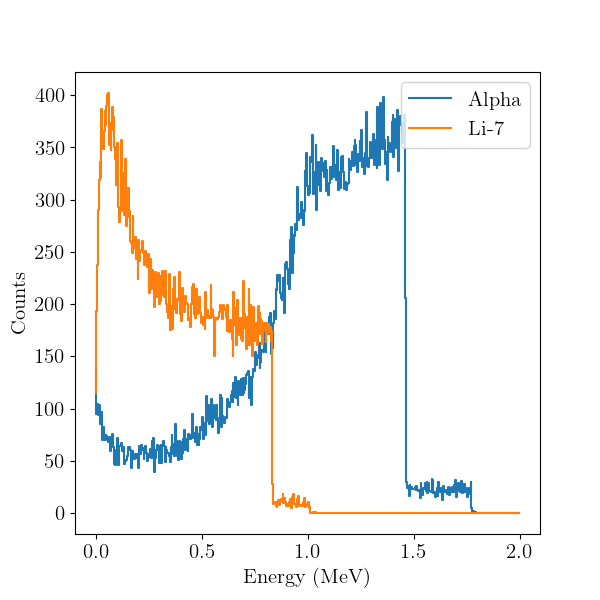}
        \caption{ }
        \label{fig:init-ion-erg}
    \end{subfigure}
    \caption{Ions entering the gas region. (a): Position and moving direction of the ions. (b): Energy spectrum of the ions.}
    \label{fig:init-ion}
\end{figure}

In the system-level model, the straw-level response, i.e., ionization and charge generation process, is not simulated. Consequently, an estimation of neutron thresholding efficiency is needed to convert the voltage threshold to deposited energy threshold. The energy deposited corresponding to a set voltage threshold can also be found through a direct comparison of simulated and measured BCS detector response. As shown in Fig.~\ref{fig:workflow}, we have developed a scheme to accurately simulate the response of the BCS detector. We first generate a list of ${}^{7}$Li/alpha ions entering the gas region by parsing the PTRAC output file from {MCNP6.2} simulation. The ion's initial energy, position, and moving direction are saved. Fig.~\ref{fig:init-ion-pos} shows the initial position and moving direction of 100 alpha/Li-7 ions, and Fig.~\ref{fig:init-ion-erg} shows the spectrum of the ions' energy when they enter the gas region. 

\begin{figure}[!htbp]
     
    \begin{subfigure}[t]{\linewidth}
        \centering
        \includegraphics[width=.9\linewidth]{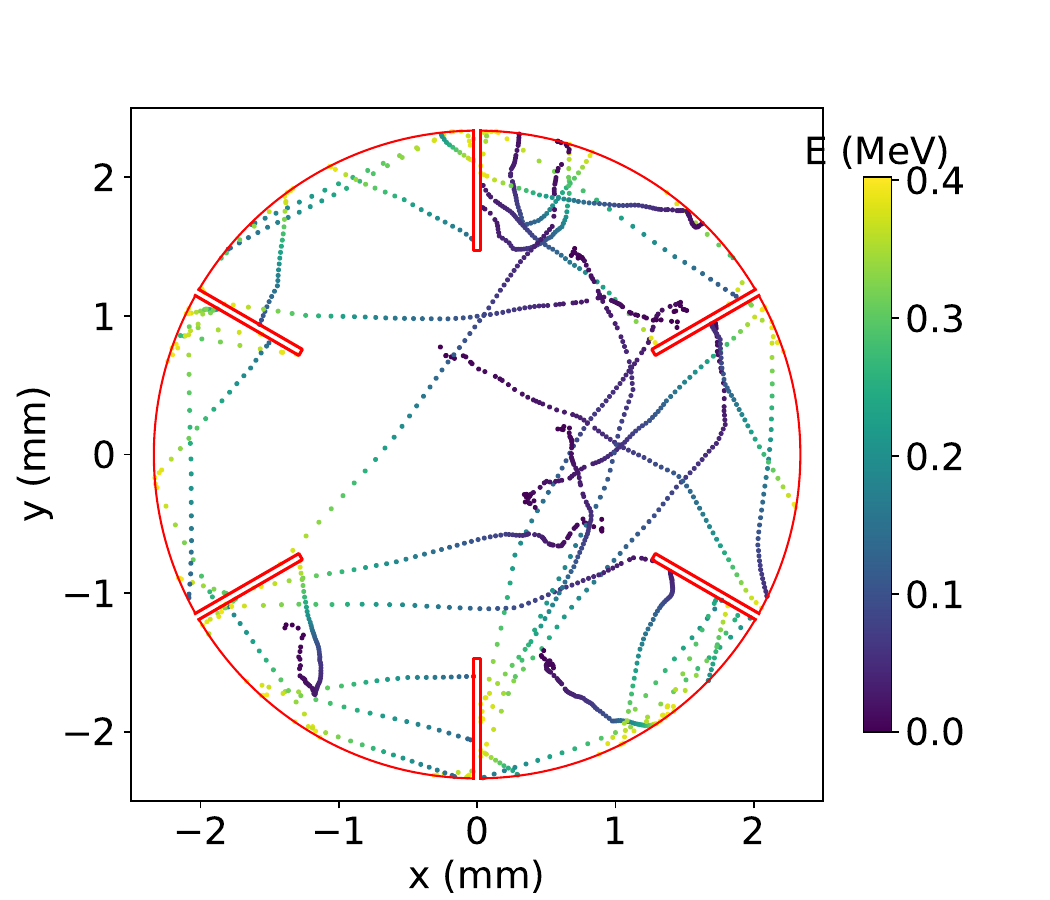}
        \caption{2D view}
        \label{fig:ion-track-2d}
    \end{subfigure}\hfil
    \begin{subfigure}[t]{\linewidth}
        \centering
        \includegraphics[width=.9\linewidth]{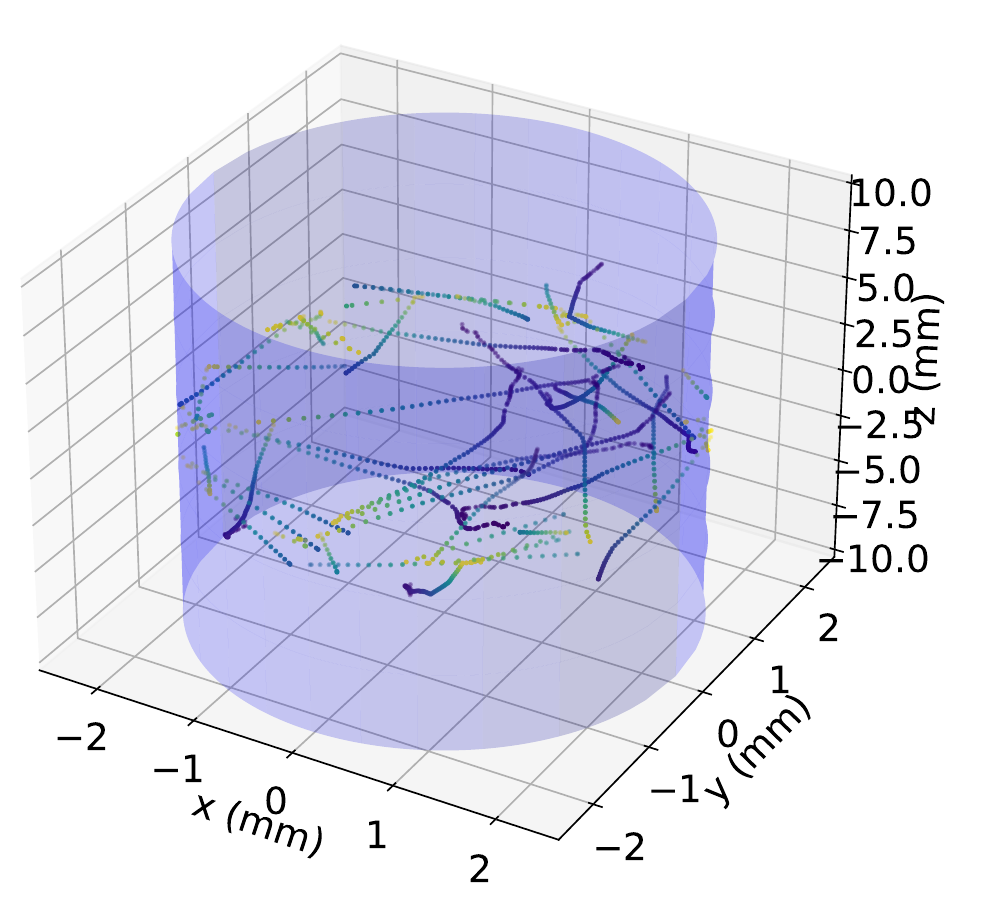}
        \caption{3D view}
        \label{fig:ion-track-3d}
    \end{subfigure}
    \caption{Tracks of 402-keV alphas in the gas region calculated by TRIM. Color represents the alpha energy at the ionizing position.}
    \label{fig:ion-track}
\end{figure}

We then use the \textit{TRIM} (Transport of Ions in Matter)~\cite{ziegler1985stopping} program to generate the ion tracks in the gas, using the automation tool \textit{pysrim}~\cite{ostrouchov2018pysrim}. TRIM simulates the track of an ion of a given energy starting at the origin and travels along the X-axis direction in a infinitely thick medium. A custom python program is developed to rotate/translate the track based on its initial position and moving direction, and terminate the track at the straw's boundary. Fig.~\ref{fig:ion-track} shows 58 alpha tracks in the gas volume. The tracks start from the wall or septa and stop at the boundary or when the ion exhausts all of its energy.

\begin{figure}[!htbp]
     
    \begin{subfigure}[t]{\linewidth}
        \centering
        \includegraphics[width=\linewidth]{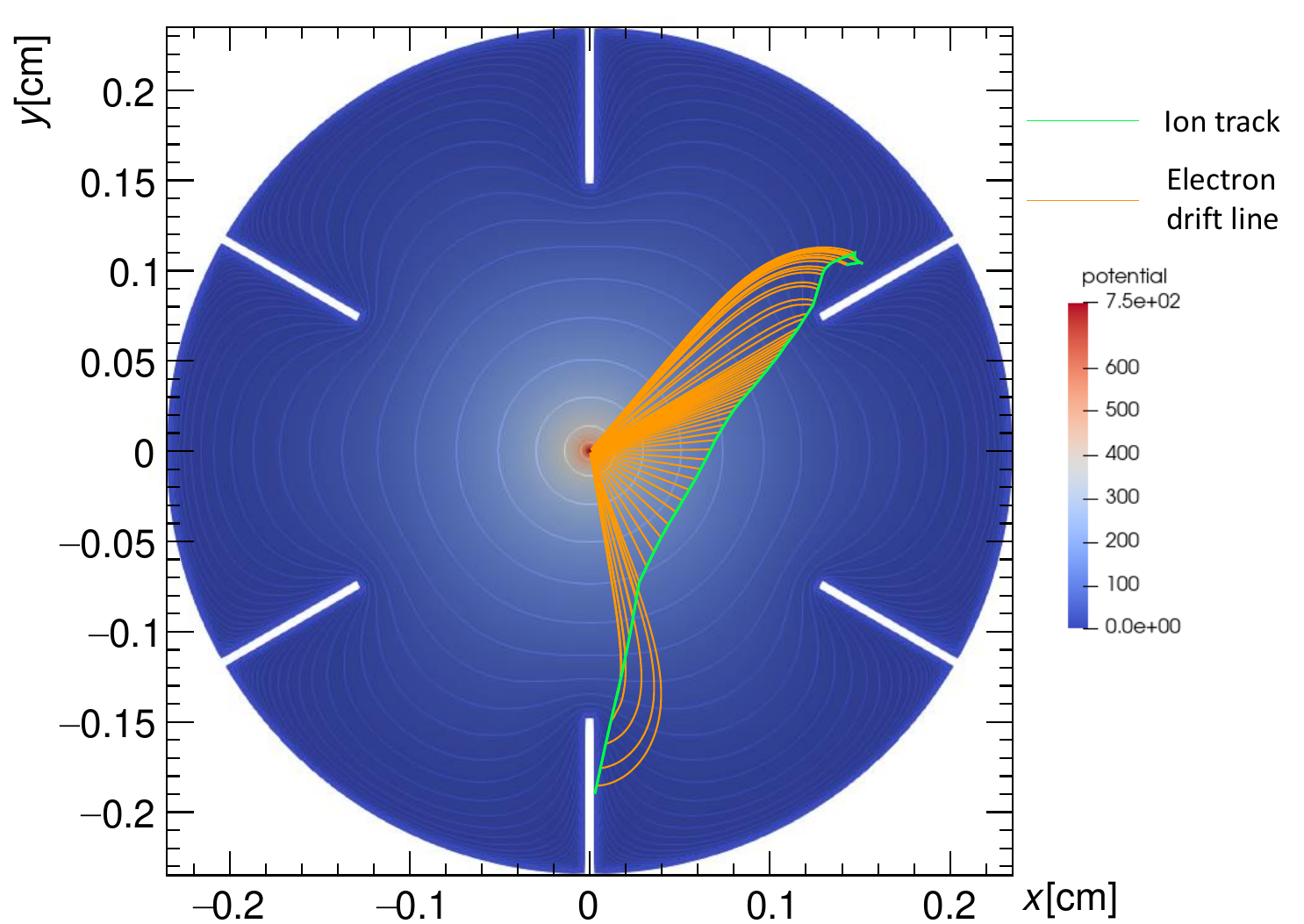}
        \caption{ }
        \label{fig:drift-lines}
    \end{subfigure}\hfil
    \begin{subfigure}[t]{\linewidth}
        \centering
        \includegraphics[width=.9\linewidth]{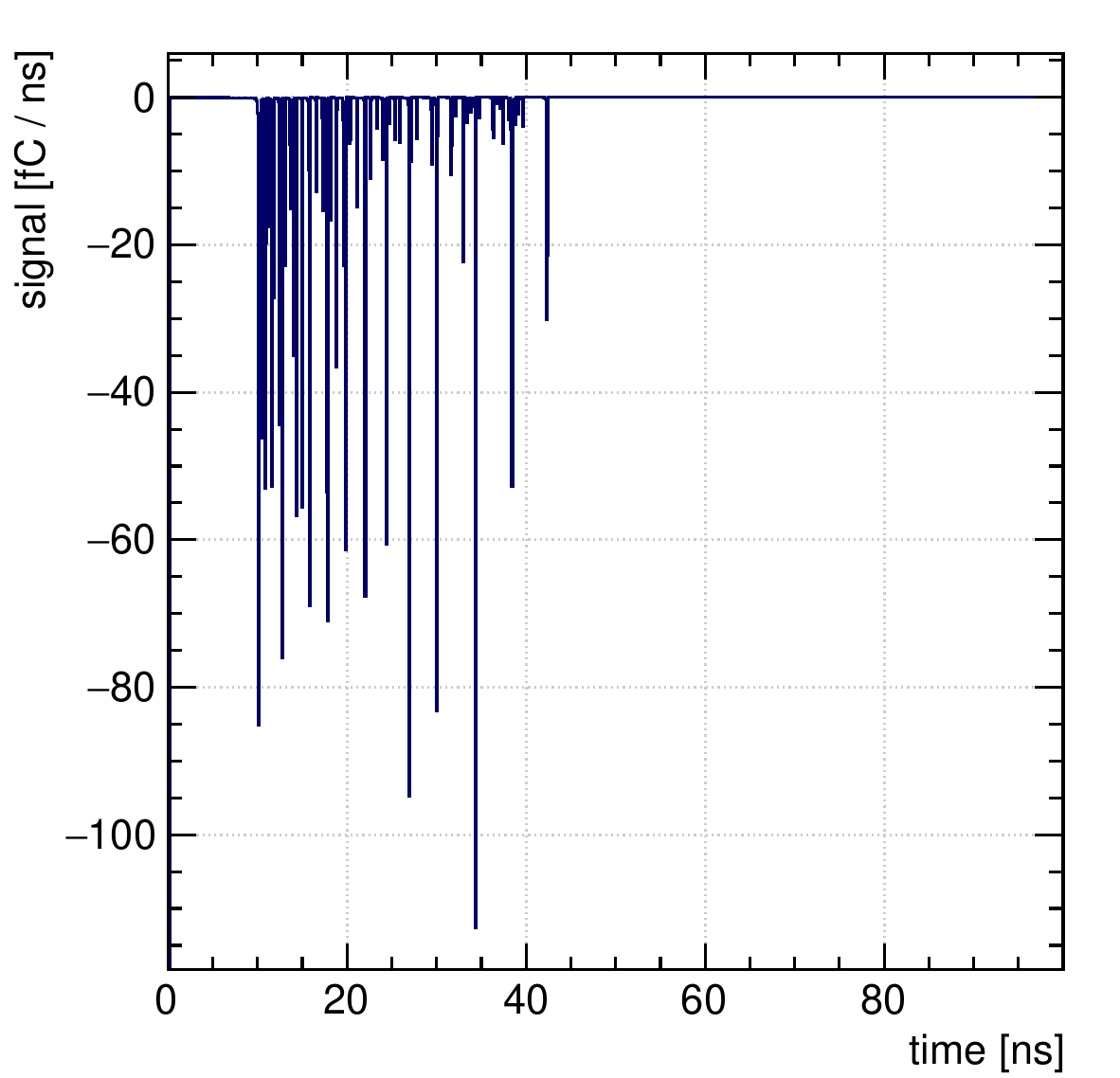}
        \caption{ }
        \label{fig:signal}
    \end{subfigure}
    \caption{(a): Drift of electrons in the electric field. (b): Current signal.}
    \label{fig:drift-line-signal}
\end{figure}

The corrected tracks are imported into \textit{Garfield++} program\cite{veenhof1993garfield}, which simulates electron avalanches in the gas volume and calculates signals induced on the wire. The electric field in the gas is solved using finite element method (FEM) tools \textit{Gmsh}~\cite{geuzaine2009gmsh} and \textit{Elmer}~\cite{ruokolainen2016elmersolver} due to the non-trivial geometry of the straw, and is imported into Garfield++. Fig.~\ref{fig:drift-lines} shows the electric field calculated by Elmer, and the drift lines of electron clusters created by a 402-keV alpha. Fig.~\ref{fig:signal} shows the current signal induced on the wire due to the drift of electrons. The total collected charge is obtained by integrating the current signal.

\section{Results}
\subsection{Validation of System-level Model}
\begin{figure}[!htbp]
     
    \begin{subfigure}[t]{\linewidth}
        \centering
        \includegraphics[width=.7\linewidth]{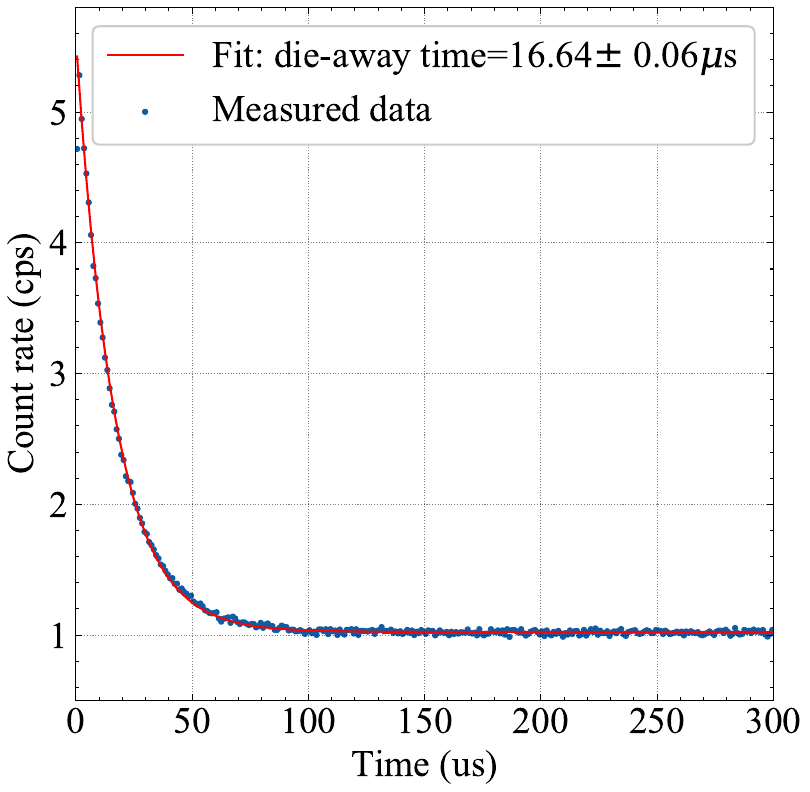}
        \caption{Experiment}
        \label{fig:rossi-alpha-exp}
    \end{subfigure}\hfil
    \begin{subfigure}[t]{\linewidth}
        \centering
        \includegraphics[width=.7\linewidth]{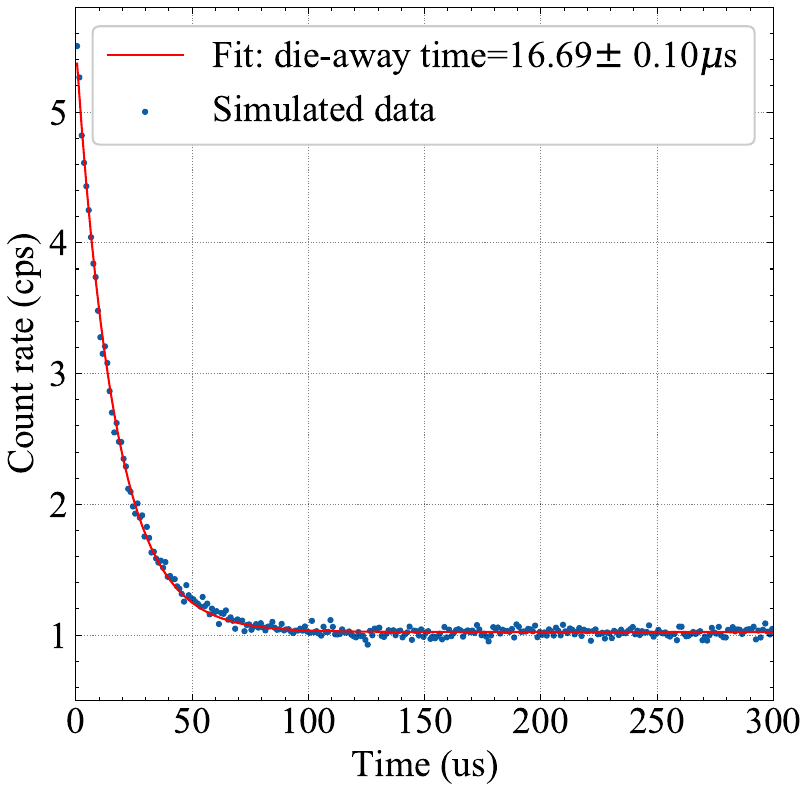}
        \caption{Simulation}
        \label{fig:rossi-alpha-sim}
    \end{subfigure}
    \caption{Comparison of the experimental and simulated Rossi-alpha distribution. Measurement time is 4500 s, and simulation time is 1000 s.}
    \label{fig:ross-alpha}
    % \vspace{-1em}
\end{figure}

We generated two lists of pulse time stamps from the PTRAC file in MCNP6 simulation and the list mode data acquired in the experiment. We plotted the Rossi-alpha distribution in Fig.~\ref{fig:ross-alpha} and fitted the data to a single exponential model with a constant offset to determine the die-away time. Table~\ref{table:sdt-comparison} shows the comparison of the experimental and simulated die-away time and coincidence rates. The relative differences are all below 0.4\%, demonstrating the high-fidelity of the NMC model.
\begin{table}[!htbp]
    \centering
     
    \caption{Comparison of the experimental and simulated single (S), double (D) count rates, and die-away time ($\tau$).}
    \label{table:sdt-comparison}
    \begin{tabular}{cccc}
    \hline
                             & S (cps) & D (cps) & $\tau$ ($\mu$s) \\ \hline
    Measurement              & 1009.8  & 59.5    & 16.64              \\
    Simulation               & 1008.9  & 59.3    & 16.69             \\
    Relative difference (\%) & \textbf{0.09}    & \textbf{0.34}    & \textbf{0.30}               \\ \hline
    \end{tabular}
\end{table}

\subsection{Validation of Straw-level model}
\begin{figure}[!htbp]
     
    \centering
    \includegraphics[width=\linewidth]{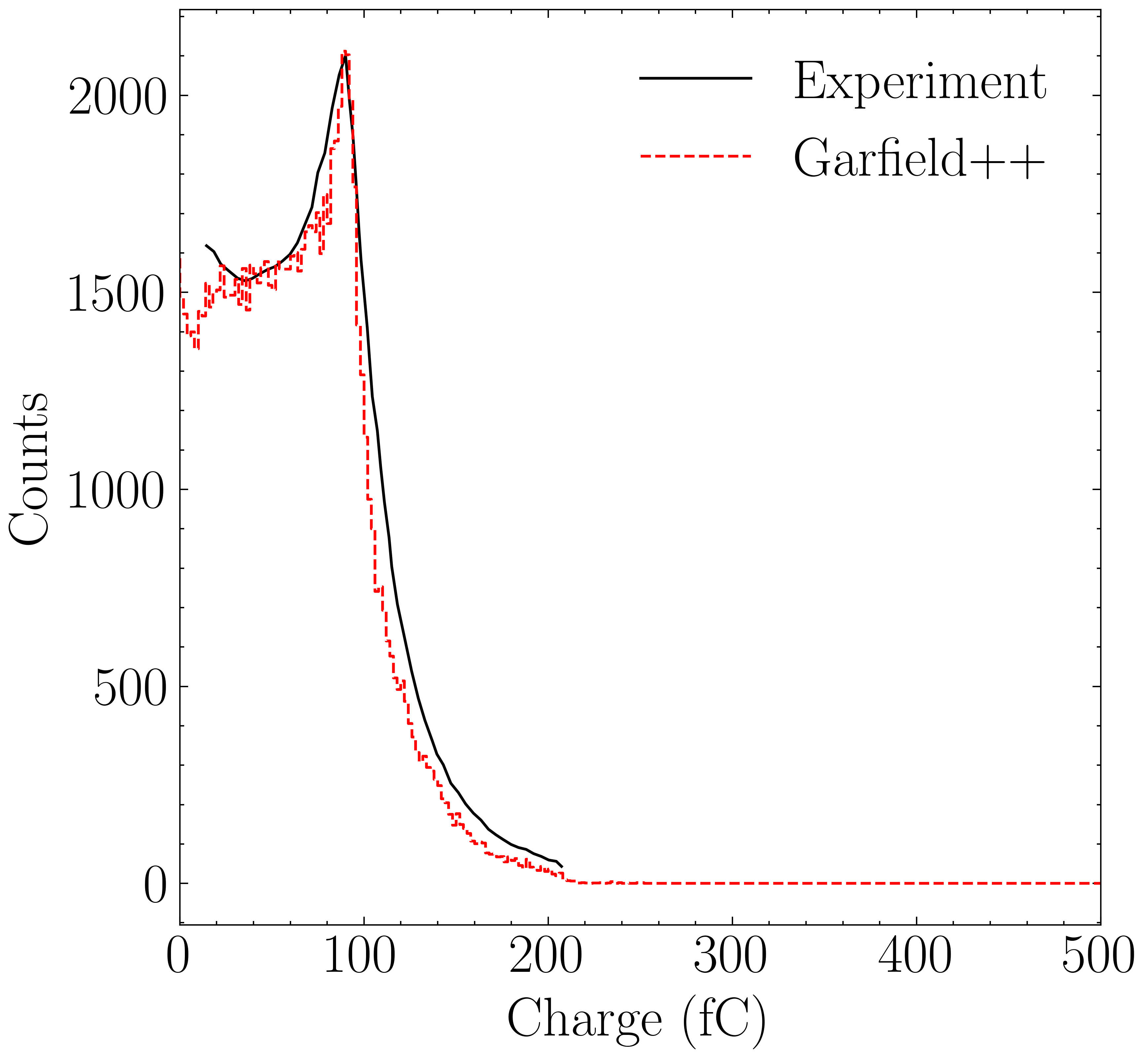}
    \caption{Comparison of the experimental and simulated spectra.}
    \label{fig:spectrum-comparison}
    % \vspace{-1em}
\end{figure}

We simulated the distribution of collected charge in a 4-mm diameter round straw using the scheme in Fig.~\ref{fig:workflow} and compared it to the experimental pulse height distribution~\cite{LACY2011359}. The bias voltage applied to the annode wire is 700~V in this case. As shown in Fig.~\ref{fig:spectrum-comparison}, the two spectra are in good agreement. 

\section{Conclusions and Future work}
In this work, we have developed a BCS-based NMC for assay of TRISO-fueled pebbles in pebble bed reactors. A system-level Monte Carlo model of the NMC has been developed. The simulated response are in good agreement with the measurement in terms of die-away time and efficiency, with $<1\%$ relative difference. A detailed simulation model of BCS was developed to simulate the charge spectrum obtained with BCS detectors. The simulated charge spectrum is in good agreement with the measurement, for a round BCS. In the future, we will account for the space charge effect and simulate the charge spectrum for pie-shaped BCS.

\bibliographystyle{IEEEtran}
\bibliography{references}
\end{document}